\documentclass[fleqn,twoside]{article}
\usepackage[headings]{espcrc2}
\usepackage{graphicx,epsfig,color}
\usepackage{array}

\readRCS
$Id: espcrc2.tex,v 1.2 2004/02/24 11:22:11 spepping Exp $
\usepackage[figuresright]{rotating}

\newcommand{\AmS}{{\protect\the\textfont2
  A\kern-.1667em\lower.5ex\hbox{M}\kern-.125emS}}
\def\twiddles#1{\mathrel{\mathop{\sim}\limits_
                        {\scriptscriptstyle {#1}}}}
\def\bea{\begin{eqnarray}}
\def\eea{\end{eqnarray}}
\def\beq{\begin{equation}}
\def\eeq{\end{equation}}
\def\mtop{m_t}
\def\ytop{y_t}
\def\mhiggs{m_H}
\def\({\left(}
\def\){\right)}
\def\[{\left[}
\def\]{\right]}

\def\as{\alpha_s}

\relax

\def    \hepph  #1 {{\tt hep-ph/#1}}
\def    \hepex  #1 {{\tt hep-ex/#1}}

\title{Finite-top-mass effects in NNLO Higgs production}

\author{Simone Marzani, \address[edi]{School of Physics and Astronomy, The University of Edinburgh, \\
Edinburgh EH9 3JZ, Scotland, UK}
Richard D.~Ball, \addressmark[edi]
Vittorio Del~Duca, \address{INFN, Laboratori Nazionali di Frascati\\
Via E. Fermi 40, I-00044 Frascati, Italy}
Stefano Forte, \address[mi]{Dipartimento di  Fisica, Universit\`a di Milano and \\
INFN, Sezione di Milano, Via Celoria 16, I-20133 Milan, Italy}
Alessandro Vicini \addressmark[mi] 
}

\begin{document}

\begin{abstract}
We construct an accurate approximation to the exact NNLO cross section for
Higgs
production in gluon-gluon fusion by matching
the dominant 
finite top mass corrections recently computed by us
to the known result in the infinite mass limit.
The
ensuing corrections to the partonic cross section are very large when
the center of mass energy of the partonic collision is much larger
than the Higgs mass, but lead to a moderate correction at the percent
level to the total Higgs production cross section
at the LHC. Our computation thus reduces the uncertainty related to these
corrections at the LHC from the percent to the per mille level.
%\vspace{1pc}
\end{abstract}

\maketitle
%\section{Introduction}
The search for the Higgs boson is one of major tasks of the
forthcoming experiments at the Large Hadron Collider (LHC) at CERN. 
The theoretical and experimental effort which has been put into Higgs
studies for LHC phenomenology is remarkable. In particular, the
determination of higher--order corrections in perturbative QCD has
been widely 
investigated. The dominant Higgs production mechanism in the
Standard Model is gluon--gluon fusion
through a top loop. The hadronic cross section can be obtained by 
convolution of the partonic cross section  with parton 
distributions $f_i(x,\mu^2)$
\bea \label{partohadr}
\sigma&&\!\!\!\!\!\!\!\!\!\!\!\!(\as;\tau_h,y_t,m_H^2)=\sigma_0(y_t)\sum_{i,j}\int_{\tau_h}^1\frac{d x_1}{x_1}\int_{\tau_h}^1 \frac{d x_2}{x_2}  \nonumber \\&&  
\!\!\!\!\!\!\!\!\!C_{i,j}\left(\as;\frac{\tau_h}{x_1 x_2},y_t\right)  f_i\left(x_1,m_H^2\right) f_j\left(x_2,m_H^2\right),
\eea
where $\sigma_0(y_t)$ is the partonic Born cross section~\cite{higgslo}
and the dimensionless variables $\tau_h$ and $\ytop$ parametrise
the hadronic center-of-mass energy and the dependence on
the top mass:
\beq
\label{ydef}
\tau_h = \frac{\mhiggs^2}{s}, \quad \ytop = \frac{\mtop^2}{\mhiggs^2} .
\eeq

The dimensionless coefficient function $C(\alpha_s;\tau,\ytop)$ 
contains the
QCD corrections. 
The NLO contribution to it was computed in~\cite{spira}
and recently confirmed in~\cite{bonciani}. 
The dominant NLO correction comes from the radiation of soft gluons,
which cannot resolve the quark loop in the ggH coupling.  Therefore,
at least at the
inclusive level, the approximation to the exact NLO result
obtained~\cite{djouadi,dawson}  by
taking the limit $\mtop\to\infty$ turns out to be very accurate.
This approximation considerably simplifies the calculation because the
ggH coupling becomes
pointlike and  the corresponding Feynman diagrams have one less
loop. 

Recently, the NNLO contribution to $C(\alpha_s;\tau,\ytop)$ has  been
computed in the $\mtop\to\infty$ limit~\cite{anastasiou}. The NNLO
result appears to be perturbatively quite stable and it should
provide a good approximation to the yet unknown exact result; it has
been widely used for precision phenomenology at the LHC~\cite{harlander}.
However, the infinite $\mtop$ approximation fails in the limit of
large partonic center--of--mass energy or equivalently \mbox{$\tau \to
  0$}.
This is due to the fact
that  the high energy behaviour of the partonic cross section is
completely different according to whether
the ggH coupling is pointlike or goes through a
quark loop, because in the latter case the quark loop effectively 
provides a form factor which softens the interaction.
Indeed,  the 
coefficient function behaves respectively as
\begin{equation} \label{hebeh}
C\twiddles{\tau \to 0}\left\{\begin{array}{c}\sum_{k=1}^\infty\as^k\ln^{2k-1}\left(\frac{1}
{\tau}\right)\quad{\rm if}\;\mtop\to\infty\\  \\
\sum_{k=1}^\infty\as^k\ln^{k-1}\left(\frac{1}
{\tau}\right)\quad {\rm for\; finite}~\mtop\\
\end{array} \right.
\end{equation}

 \begin{figure}
 \begin{center}
 \vspace{0.5cm}
\includegraphics[width=.45\textwidth]{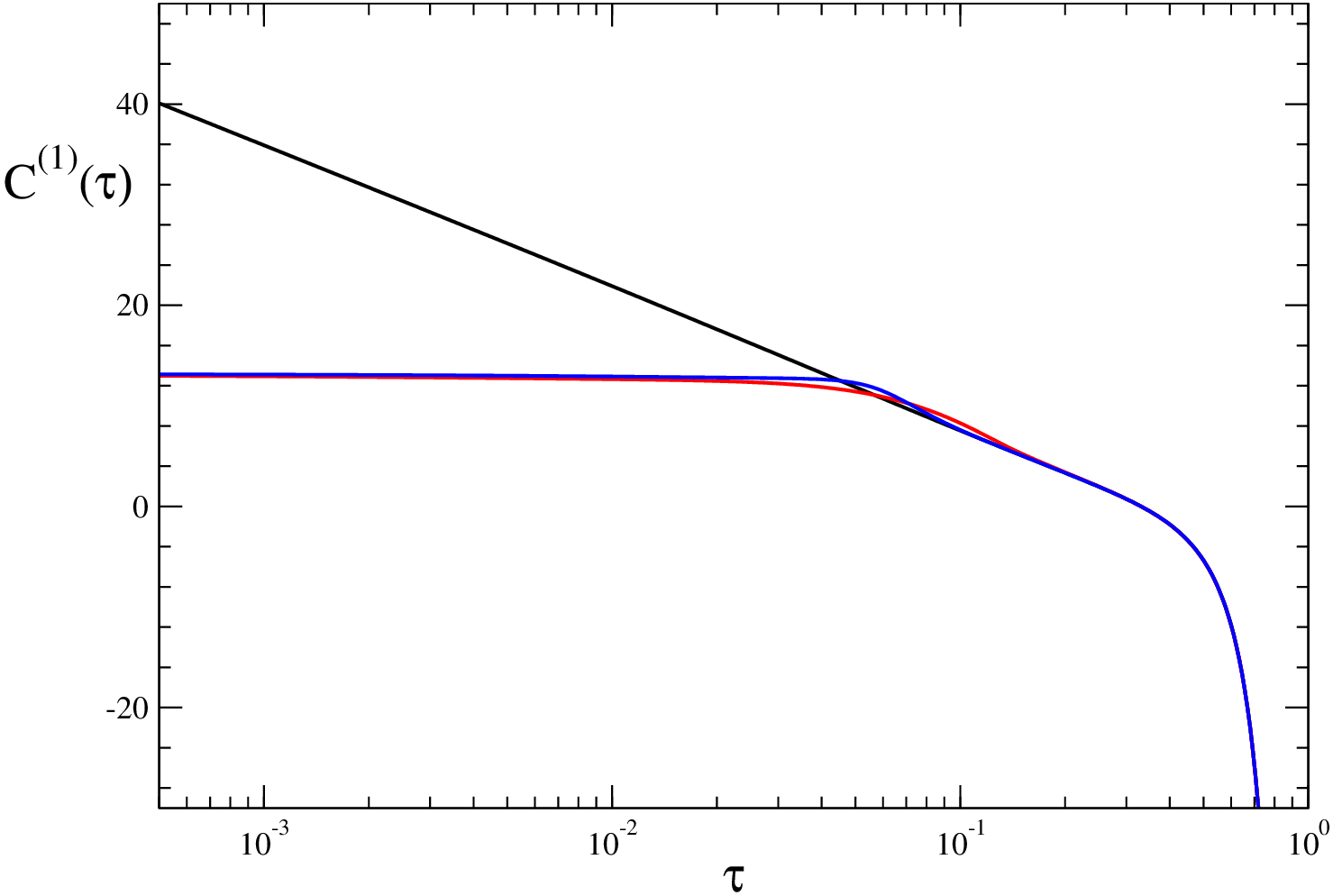} 
 \caption{The NLO coefficient function for \mbox{$m_H=130$~GeV}. The
   red curve corresponds to the exact case, the black one to
   \mbox{$m_t \to \infty$} and the blue one to the
   approximation Eq.~(\ref{appr}) with $\tau_0=0.057$ and $\omega=1/100$.}\label{fig:nlo130} 
 \end{center}
\vskip-.8cm
\end{figure}
Equation~(\ref{hebeh}) shows that the difference at high energy between the
exact and approximate behaviour is  larger at higher orders, so
one might expect the relative accuracy of the infinite $\mtop$
approximation to become accordingly worse.
In Ref.~\cite{bdfmv} we have recently 
computed the leading high energy logarithms Eq.~(\ref{hebeh})
at finite $m_t$, using the techniques of  high-energy (or \mbox{$k_T$)
factorization}~\cite{catani} (the corresponding coefficients in the
$\mtop\to\infty$ limit had been previously computed in Ref.~\cite{hautmann}).
We can use this result to construct an improvement of the NNLO
result~\cite{anastasiou}, 
 by replacing its spurious double logarithmic
growth with the correct high energy behaviour, Eq.~(\ref{hebeh}).

This construction requires a suitable matching procedure, and it 
gives us an approximation to the exact NNLO result. We shall first
perform this improvement on the NLO contribution, where the exact result is
known: it turns out to  give an approximation to the exact NLO
partonic cross section which is everywhere accurate to better than
1~\% for any value of the Higgs mass, thus leading to an approximation
to the total cross section which is accurate to the level of 0.05\%. 
We shall than construct a
similar improvement of the NNLO term. This
improvement changes the total cross section computed up to NNLO 
by an amount which
varies between 0.2\% for light Higgs ($m_H\sim 130$~GeV) to 1\% for
heavy Higgs ($m_H\sim 280$~GeV). This is thus the size of the error
which is made if the cross section is computed 
at NNLO using the
approximate $\mtop\to\infty$ result. By varying the matching prescription,
we estimate that the ambiguity on this result is at the level of the
per mille.

  \begin{figure}
 \begin{center}
 \vspace{.5cm}
\includegraphics[width=.45\textwidth]{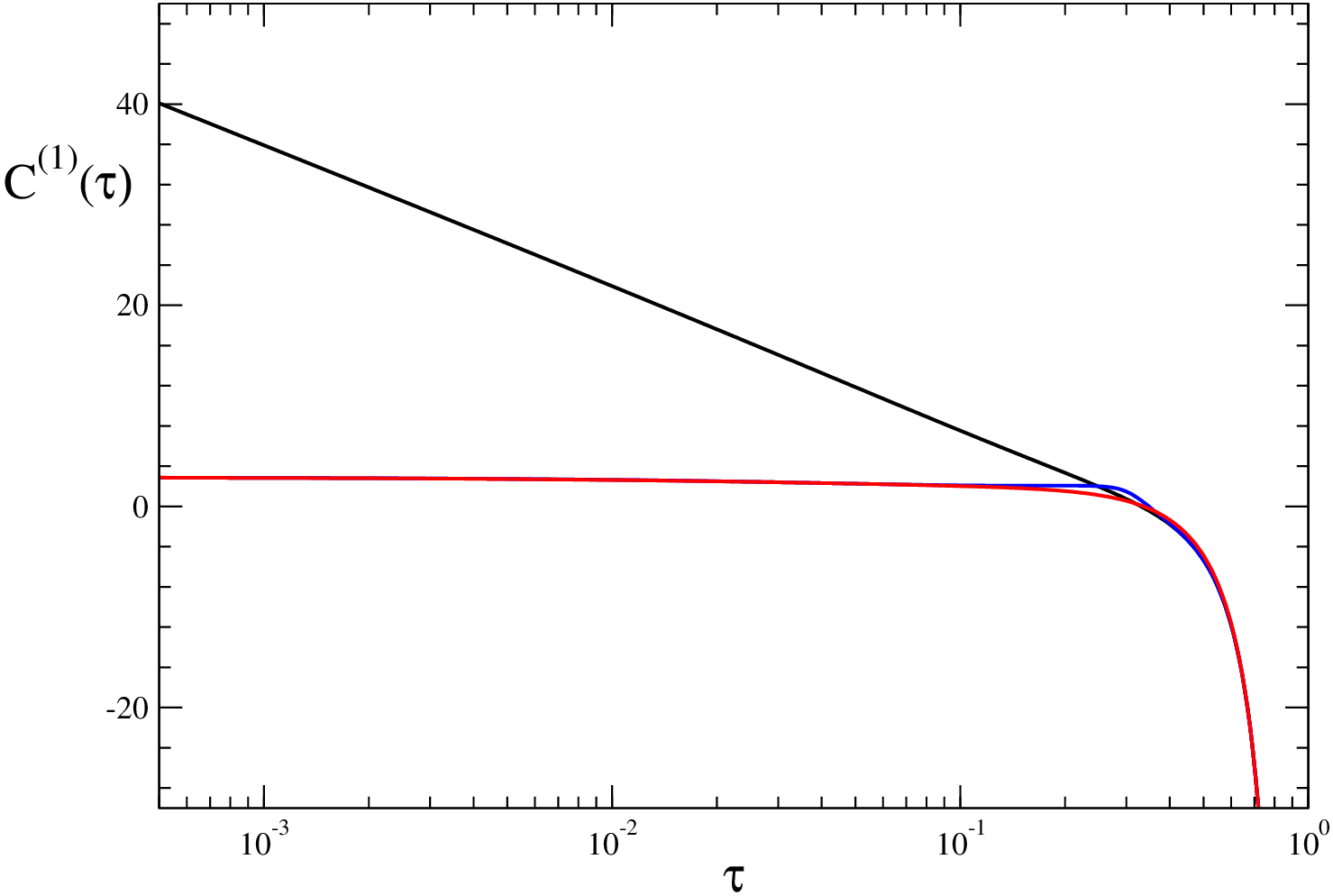} 
 \caption{Same as Fig.~\ref{fig:nlo130}, but with \mbox{$m_H=280$~GeV}
   (here $\tau_0=0.315$ and $\omega=1/20$).}\label{fig:nlo280}
 \end{center}
\vskip-.4cm
\end{figure}
\begin{figure}
\goodbreak
 \begin{center}
 \vspace{0.8cm}
\includegraphics[width=.45\textwidth]{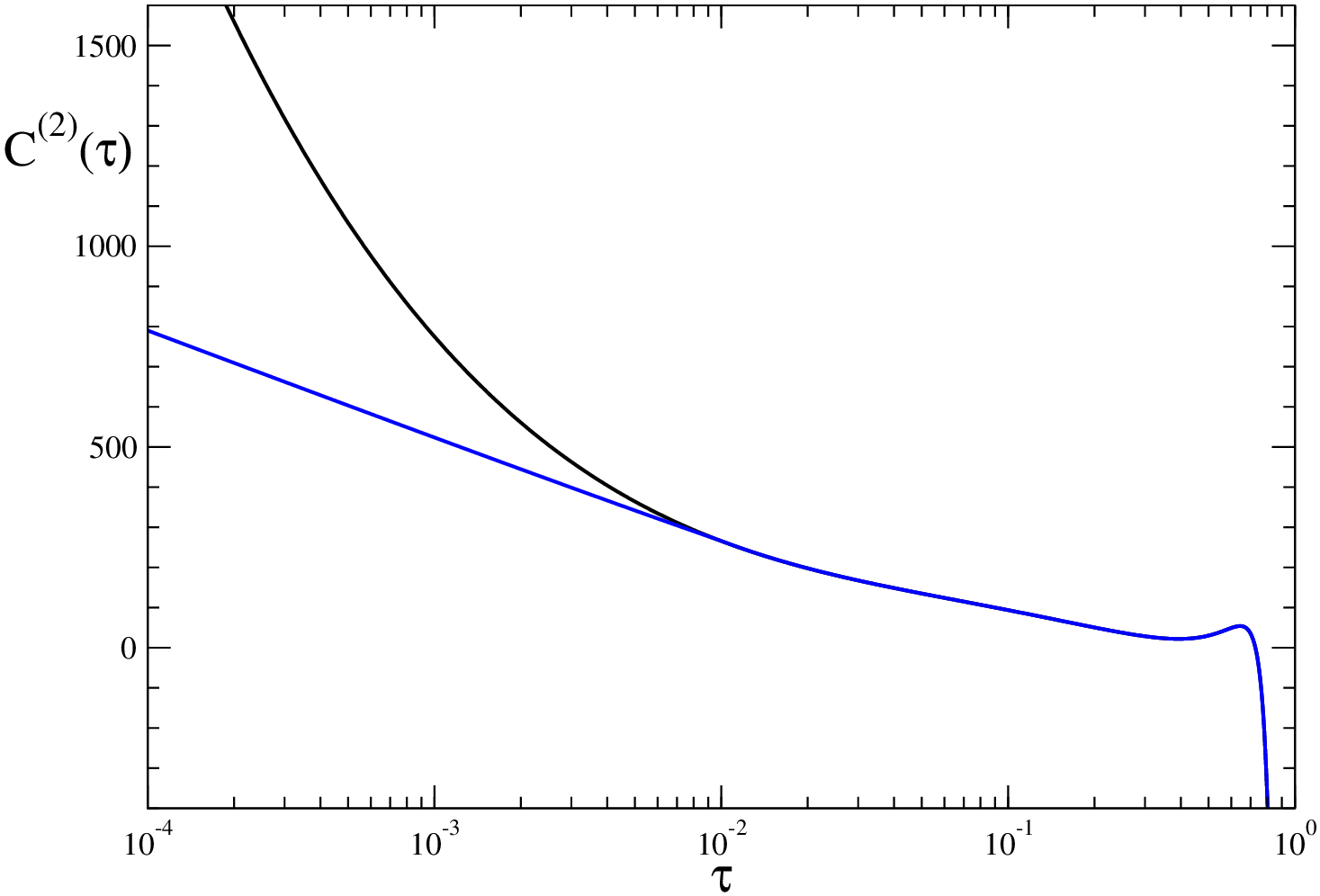}
\caption{The NNLO coefficient function for $m_H=130$~GeV. The black
  curve corrdesponds to \mbox{$m_t \to \infty$} and the blue one to the
  approximation Eq.~(\ref{appr}) with $\tau_0=0.011$ and
  $\omega=1/100$. 
The exact case is not known at
  NNLO.}\label{fig:nnlo130} 
 \end{center}
\vskip-.7cm
\end{figure}
%\section{Improvement of the $m_t \to \infty$ cross section}
The perturbative expansion of the
coefficient function in the gluon-gluon channel is 
\bea
&&C(\as;\tau,\ytop)=\delta(1-\tau)+\frac{\as}{\pi}
C^{(1)}(\tau,\ytop)\nonumber \\  &&\qquad\qquad+\left(\frac{
    \as}{\pi}\right)^2
C^{(2)}(\tau,\ytop)+ \mathcal{O}\left(\as^3 \right). 
\eea
The leading high energy behaviour of the NLO and NNLO contributions is
given by
\bea \label{smalltau}
 &&\!\!\!\!\!\! C^{(1)}(\tau,\ytop) = {\mathcal
    C}_1^{(1)}(\ytop) C_A + O(\tau)   \\
 &&\!\!\!\!\!\!C^{(2)}(\tau,\ytop)= -{\mathcal C}_2^{(2)}(\ytop) C_A^2 \ln \tau + {\mathcal C}_1^{(2)}(\ytop) 
  + O(\tau).  \nonumber 
\eea %The coefficient ${\mathcal
    %C}^{(1)}(\ytop)$  is the small $\tau$ limit of the full NLO computation and 
    The leading NNLO coefficient  ${\mathcal
    C}_2^{(2)}(\ytop)$  was computed in Ref.~\cite{bdfmv}, while  the
  subleading NNLO coefficient ${\mathcal
    C}_1^{(2)}(\ytop)$ is unknown. 

The approximate
pointlike determination of the coefficient function can be improved 
by subtracting
its spurious small~$\tau$ growth and replacing it with the exact
behaviour:
 \bea \label{appr}
&&C^{\rm app.}(\tau,\ytop)= C(\tau,\infty)+T(\tau,\tau_0)
\times\nonumber \\ &&\qquad\left[C(\tau,\ytop) 
-\lim_{\tau \to 0}C_0(\tau,\infty) \right] ,
 \eea
where $C_0(\tau,\infty)$ is the sum of contributions to the infinite $m_t$
result $C(\tau,\infty)$ which do not vanish when $\tau\to0$, as given
at NLO and NNLO by the terms listed in Eq.~(\ref{smalltau}).
Also,
$T(\tau,\tau_0)$ is a matching function, which is introduced in
  order to tune the point $\tau_0$ where the small~$\tau$ behaviour sets in.
  We choose
 \beq\label{match}
 T(\tau,\tau_0)= \frac{1}{2} \left[1+ {\rm tanh}\left(\frac{\tau_0-\tau}{\omega}\right) \right],
 \eeq
which in the limit of  vanishing
 width $\omega$ becomes the step function:
$\lim_{\omega \to 0}
 T(\tau,\tau_0)= \Theta(\tau-\tau_0)$. 

 At NLO, because the exact asymptotic behaviour Eq.~(\ref{smalltau}) is
 the constant ${\mathcal C}_1^{(1)}(\ytop)$, the 
 matching point $\tau_0$ is naturally determined as the
 value of $\tau$ where the pointlike
  approximation equals this constant. It is clear from
  Fig.~\ref{fig:nlo130} that this choice leads 
to an excellent  approximation to the exact result:
in fact, it is accurate to better than 1\% for all
$\tau$.

 \begin{figure}
 \begin{center}
 \vspace{0.5cm}
\includegraphics[width=.45\textwidth]{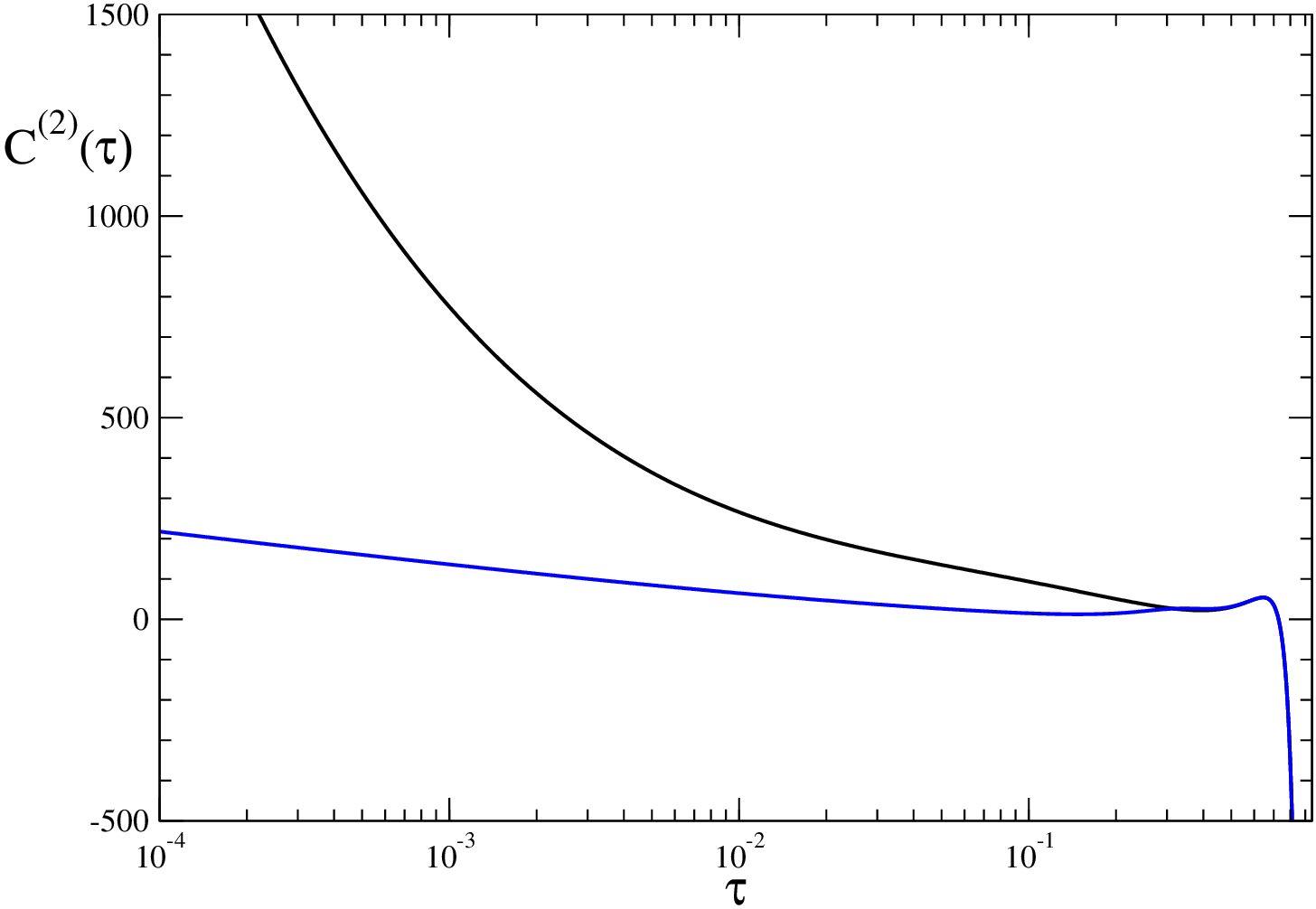}
\caption{Same as Fig.~\ref{fig:nnlo130}, but with \mbox{$m_H=280$~GeV}
  (here $\tau_0=0.317$ and  $\omega=1/20$).}\label{fig:nnlo280}
 \end{center}
\vskip-.5cm
\end{figure}
At NNLO the exact asymptotic behaviour Eq.~(\ref{smalltau}) is a
linear rise in $\ln \tau$. Hence, reasoning as at NLO, we are led to
choose 
 $\tau_0$ as the point where the log derivative of the $m_t \to
 \infty$ curve
matches the asymptotic value ${\mathcal C}_2^{(2)}(\ytop)$:
 \begin{equation} \label{continuityder}
  \frac{d}{d \ln \tau} C^{(2)}(\tau,\infty) \Big|_{\tau=\tau_0} =
  -9\mathcal{C}^{(2)}(\ytop). 
 \end{equation}
This does not fix completely the approximate result at NNLO however,
because the subleading constant ${\mathcal C}_1^{(2)}(\ytop)$ is
unknown. We fix it by requiring that the approximate curve
Eq.~(\ref{appr}) be
continuous at $\tau=\tau_0$ even when the matching function
$T(\tau,\tau_0)=\Theta(\tau-\tau_0)$.
The NNLO approximation determined thus  is compared to the exact result in
Figs.~\ref{fig:nnlo130}, \ref{fig:nnlo280} for heavy and light Higgs,
respectively. 
A more conservative matching might consist instead of taking  for
$\tau_0$  the value found at NLO, and then determining
again
the subleading constant by continuity. The result found in this way is
actually very close to the previous one, and in fact indistinguishable
from it in the case of Fig.~\ref{fig:nnlo280}.

Let us now turn to the inclusive cross section.
We define a $K$ factor 
%NLO and NNLO $\kappa$ factors by letting
\beq\label{kfactdef}
K(\tau_h;\ytop)\equiv
\frac{\sigma_{gg}(\tau_h,\ytop,\mhiggs^2)}{\sigma^{0}_{gg}(\tau_h;\ytop,\mhiggs^2)},\eeq where $\sigma^{0}_{gg}$ is the
LO cross-section Eq.~(\ref{partohadr}), computed
with LO parton distributions and LO coupling constant. The value
of the NLO and NNLO $K$ factors, determined using the
MRST2002~\cite{mrst2002} gluon distribution in
Eq.~(\ref{partohadr}) are given in Table~\ref{tabkfact}, at the LHC
center-of-mass energy $s=14$~TeV. In the table the pointlike,
exact and approximate cases are shown.

\begin{table}[t!]
  \begin{center}
 \begin{tabular}{|c|c|c|}
 \hline
   & $K^{\rm NLO}$ & $K^{\rm NNLO}$ \\
 \hline
\multicolumn{3}{|c|}{$\mhiggs=130$~GeV}\\
\hline
 pointlike   &  1.800 &  2.140   \\
 exact   &  1.797 &  n.a.  \\
appr.   &  1.796 & 2.136     \\

 \hline
\multicolumn{3}{|c|}{$\mhiggs=280$~GeV}\\
\hline
 pointlike   &  1.976 & 2.420   \\
 exact   & 1.958  &  n.a.   \\
appr.    & 1.959 & 2.394   \\

 \hline
  \end{tabular}\\
   \caption{The NLO and NNLO $K$ factors
 Eq.~(\ref{kfactdef}), computed with
     center-of-mass energy $s=14$~TeV. %and  $\mtop\to\infty$.% denoted with
     %pointlike, or $\mtop=170.9$~GeV, denoted with  exact or
     %approximate. 
}\label{tabkfact}
   \end{center}
\vskip-.9cm
  \end{table}
 At NLO the discrepancy between the infinite top mass approximation
 and the exact result is tiny, less than $1\%$ even for a fairly heavy
 Higgs. If the improved (approximate) NLO result is used, this discrepancy is
 reduced by a factor three.
 At NNLO the inclusion of the correct small $\tau$ dependence of the
 partonic coefficient function changes the $K$ factor by  an amount 
which varies between $0.3 \%$ for
 \mbox{$m_H=130$~GeV} and $1\%$ for \mbox{$m_H=280$~GeV}. If we
 modify the matching prescription by using the NLO value of $\tau_0$
 also at NNLO the approximate NNLO results of table~1 change by
 0.1 \%. We can take this as the error which is made by use of the
 infinite $\mtop$ NNLO formula (there is also a dependence on $\mtop$
in the contribution to $C^{(2)}$ which is proportional to
 $\delta(1-\tau)$, but at NNLO  this contribution is relatively small,
 unlike at NLO). Dominant uncertainties on the total 
 Higgs cros-section are typically at the percent level~\cite{harlander}.
Varying $\omega$ in the
 matching function Eq.~(\ref{match})
between $1/20$ and $1/100$  the NNLO  results change by about
0.1~\%. 
We conclude that use of the improved approximate NNLO reduces the
error due to finite mass terms at NNLO to the per mille level.
This may be relevant in view of recent progress on the computation of
electroweak corrections to this process~\cite{ewcorr}.

Finally, we observe that 
less inclusive quantities which depend on the $\tau$ shape of the
partonic cross section can  be rather more sensitive to finite--mass
effects, in particular if they probe the small~$\tau$ 
tail of a coefficient function. An interesting case in point, which
deserves further
investigation, is the Higgs rapidity distribution~\cite{nnlohiggsdy}.

\end{document}